\documentclass[fleqn,usenatbib]{mnras}
\usepackage{newtxtext,newtxmath}

\usepackage[T1]{fontenc}
\DeclareRobustCommand{\VAN}[3]{#2}
\let\VANthebibliography\thebibliography
\def\thebibliography{\DeclareRobustCommand{\VAN}[3]{##3}\VANthebibliography}
\usepackage{graphicx}	
\usepackage{amsmath}	

\title[Independence of Kink Oscillation Periods of Noise]{Do Periods of Decayless Kink Oscillations of Solar Coronal Loops Depend on Noise?}
 
\author[V.M.~Nakariakov et al.]{
Valery M. Nakariakov,$^{1,2}$\thanks{E-mail: V.Nakariakov@warwick.ac.uk}
Dmitrii Y. Kolotkov,$^{1}$
Sihui Zhong$^{1}$
\\
$^{1}$ Centre for Fusion, Space and Astrophysics, Department of Physics, University of Warwick, Coventry CV4 7AL, UK\\
$^{2}$ Centro de Investigacion en Astronom\'ia, Universidad Bernardo O'Higgins, Avenida Viel 1497, Santiago, Chile\\}

\date{Accepted XXX. Received YYY; in original form ZZZ}

\pubyear{2022}

\begin{document}
\label{firstpage}
\pagerange{\pageref{firstpage}--\pageref{lastpage}}
\maketitle

\begin{abstract}
Decayless kink oscillations of solar coronal loops are studied in terms of a low-dimensional model based on a randomly driven Rayleigh oscillator with coefficients experiencing random fluctuations. The model considers kink oscillations as natural modes of coronal loops, decaying by linear resonant absorption. The damping is counteracted by random motions of the loop footpoints and the interaction of the loop with external quasi-steady flows with random fluctuations. In other words, the model combines the self-oscillatory and randomly driven mechanisms for the decayless behaviour. The random signals are taken to be of the stationary red noise nature. In the noiseless case, the model has an asymptotically stationary oscillatory solution, i.e., a kink self-oscillation. It is established that the kink oscillation period is practically independent of noise. This finding justifies the seismological estimations of the kink and Alfv\'en speeds and the magnetic field in an oscillating loop by kink oscillations, based on the observed oscillation period. The oscillatory patterns are found to be almost harmonic. Noisy fluctuations of external flows modulate the amplitude of the almost monochromatic oscillatory pattern symmetrically, while random motions of the loop footpoints cause antisymmetric amplitude modulation. Such modulations are also consistent with the observed behaviour. 
\end{abstract}

\begin{keywords}
Sun: corona -- Sun: oscillations -- waves
\end{keywords}



\section{Introduction}

Kink oscillations of solar coronal loops are subject to intensive ongoing studies \citep[see][for a recent comprehensive review]{2021SSRv..217...73N}. Oscillations in this mode are characterised by periodic transverse displacements of the loop axis, and are, in the long-wavelength limit, weakly-compressive. Kink oscillations are confidently detected with high-resolution EUV imaging telescopes as periodic displacements of loops in the plane of the sky, and interpreted as standing waves \citep[e.g.][]{1999ApJ...520..880A, 2011ApJ...736..102A}. In addition, Doppler shift oscillations of coronal emission lines and periodic variations of the intensity of the microwave emission produced in a solar flare have been interpreted as kink oscillations too \citep[see][respectively]{2012ApJ...759..144T,  2013SoPh..284..559K}. 

Standing kink oscillations are observed in two regimes, a rapidly-decaying large amplitude one \citep[e.g.][]{1999Sci...285..862N}, and a decayless low-amplitude one \citep{2012ApJ...751L..27W}. The oscillation periods scale linearly with the length of the oscillating loop \citep{2002ApJ...576L.153O, 2016A&A...585A.137G, 2015A&A...583A.136A, 2019ApJS..241...31N}, indicating that the oscillations are natural (eigen) modes. In the decaying regime, the linear scaling of the damping time with oscillation period is consistent with the oscillation damping due to linear coupling between the kink mode and highly localised torsional oscillation, i.e. the phenomenon of resonant absorption \citep[e.g.][]{2002ApJ...577..475R, 2006RSPTA.364..433G, 2010ApJ...711..990P}. The existence of the exponential and Gaussian decaying patterns strengthens this interpretation \citep[e.g.][]{2013A&A...551A..40P}. However, the empirically determined dependence of the oscillation quality factor on the amplitude \citep{2016A&A...590L...5G} suggests that the actual process may be more complicated. In particular, the oscillating loop could be subject to wave-induced Kelvin--Helmholtz instability \citep[e.g.][]{2016A&A...595A..81M, 2017ApJ...836..219A}. Decaying kink oscillations are usually excited by a displacement of the loop from equilibrium by a low-coronal eruption \citep{2015A&A...577A...4Z}. 

\begin{figure}
	\includegraphics[width=\columnwidth]{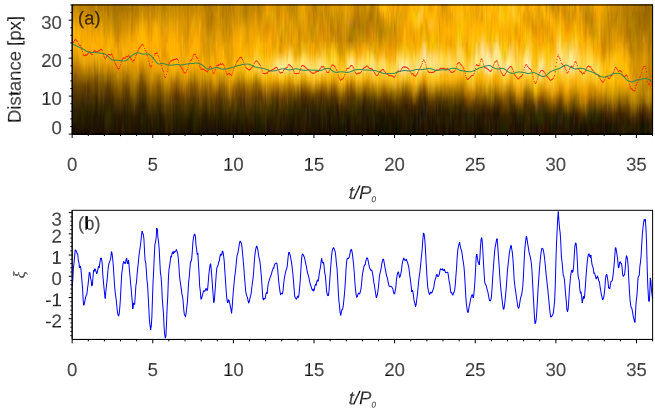}
    \caption{Example of a decayless kink oscillation of a coronal loop, detected from 21:24 UT on 24th June 2021 to 00:33 UT on 25th June 2021 at AIA 171\,\AA. Panel (a): Time--distance map overplotted with oscillation displacement amplitudes (red) and its trend (green). The trend is obtained by smoothing the original time series with window longer than the oscillation period. Panel (b): The detrended oscillatory pattern (blue). The oscillation amplitude is magnified by a factor of 4, and normalised to the standard deviation. The time is measured in the units of the mean instantaneous oscillation period $P_0$. See \citet{2022MNRAS.513.1834Z} for more details.
    }
    \label{fig:fig1}
\end{figure}

In the other, decayless regime, kink oscillations are persistent \citep[e.g.][]{2012ApJ...751L..27W}, and are seen to last up to several tens of cycles, see Fig.~\ref{fig:fig1} \citep{2022MNRAS.513.1834Z}. Typically, amplitudes of  decayless oscillations are about an order of magnitude smaller than of decaying oscillations \citep{2015A&A...583A.136A}, but large amplitude decayless oscillations have been detected too \citep{2021A&A...652L...3M}.  Oscillations with similar properties have recently been detected in coronal bright points \citep{2022ApJ...930...55G}. As both decaying and decayless oscillatory regimes are observed in different time intervals in the same loop \citep{2013A&A...552A..57N}, the effect of resonant absorption should occur in both the regimes. It raises a question of which mechanism compensates the energy losses to torsional oscillations in the decayless regime. As the distribution of amplitudes over oscillation periods rules out resonant excitation by a periodic force, the models of randomly driven oscillations \citep{2020A&A...633L...8A, 2021MNRAS.501.3017R, 2021SoPh..296..124R} and self-oscillations \citep{2016A&A...591L...5N, 2020ApJ...897L..35K} have been proposed. In the former case, the energy comes from random motions of the loop's footpoints, e.g., by the granulation, while in the latter case from the interaction of the loop with quasi-steady, on the time scale of the oscillation period, external flows. The kink self-oscillation is analogous to the oscillating violin string excited by a bow moving across the string via negative friction \citep[e.g.][]{2013PhR...525..167J}.  Alternative suggestions have linked decayless oscillatory patterns with the development of the Kelvin--Helmholtz instability \citep{2016ApJ...830L..22A} and a pattern of interference fringes \citep{2014ApJ...784..103H}.

Our interest in decayless kink oscillations is connected with their role in the energy balance in the magnetically closed corona \citep{2019FrASS...6...38K, 2020SSRv..216..140V, 2021ApJ...908..233S}, and also with the unique opportunity to perform seismological diagnostics of coronal plasmas during the quiet periods of the solar activity \citep{2019ApJ...884L..40A,2020ApJ...894L..23M}. In particular, the seismological estimation of the absolute value of the magnetic field in oscillating loops is based on the assumption that the oscillation period coincides with the period of a standing kink wave  \citep{2001A&A...372L..53N}.  On one hand this assumption is consistent with the empirically established linear dependence of the oscillation period on the loop length \citep{2015A&A...583A.136A}. Moreover, numerical modelling shows that the variation of the perpendicular profile of the fast speed does not change the period for more than a few percent \citep[][]{2019FrASS...6...22P}. However, it is not clear how sensitive this assumption is to the presence of noise. This question is important, as the noise is an intrinsic feature of the oscillation excitation mechanism by random footpoint motions, and also could be present in the self-oscillatory model via the noisy character of the external flow. 

In this letter we combine the randomly-driven and self-sustained oscillation models to demonstrate that the kink oscillation period is almost insensitive to the noisy dynamics around the oscillating loop and at its footpoints. In Sec.~\ref{SecM} we describe the model, in Sec.~\ref{SecR} we present the results, and in Sec.~\ref{SecD} we summarise our conclusions.

\section{The model}
\label{SecM}

Following \citet{2016A&A...591L...5N}, we consider decayless kink oscillations in terms of a zero-dimensional model based on a simple harmonic oscillator, 
\begin{equation}
\label{goveEq1}
\ddot{\xi} +  \delta \dot{\xi} + \Omega_\mathrm{k}^2 \xi = F(v_0 - \dot{\xi}) + N(t), 
\end{equation}
where $\xi$ is the oscillation amplitude, $\delta$ is a constant which describes damping by, e.g., resonant absorption, $\Omega_\mathrm{k}$ is the natural frequency of a standing kink wave, i.e., $\Omega_\mathrm{k} = \pi C_\mathrm{k}/L$ with $C_\mathrm{k}$ being the kink speed; $F$ describes the interaction of the loop with an external flow $v_0$, and $N$ accounts for random motions at the footpoints. Thus, on the left hand side we gather the terms which describe a kink oscillation as a decaying eigen mode of the loop, while on the right hand side there are terms responsible for the oscillation excitation by quasi-steady external flows via the negative friction and random motion of footpoints. 
Performing the Taylor expansion of the function $F$, we obtain a generalised driven Rayleigh oscillator equation,
\begin{equation}
\label{goveEq2}
\ddot{\xi} +  \left[ (\delta-\delta_v) +\alpha \left(\dot{\xi}\right)^2\right]\dot{\xi} + \Omega_\mathrm{k}^2 \xi = N(t). 
\end{equation}
In the following, we assume that the coefficients of the Taylor expansion, $\delta_v$ and $\alpha$, fluctuate randomly around the mean values $\delta_{v0}$ and $\alpha_0$, i.e,  $\delta_v = \delta_{v0} + \delta_{v1}\eta_v(t)$ and $\alpha = \alpha_0 + \alpha_1\eta_\alpha(t)$, and $N = N_0 \eta_\mathrm{N}(t)$. The functions $\eta_v(t)$, $\eta_\alpha(t)$ and $\eta_\mathrm{N}(t)$ are stationary dimensionless red noises with zero means and the standard deviations of unity. The parameters $\delta_{v1}$, $\alpha_1$ and $N_0$ are amplitudes of those noises, respectively. This kind of noise is consistent with empirically determined dynamics of the corona \citep[e.g.][]{2015ApJ...798....1I, 2016A&A...592A.153K, 2019ApJ...886L..25Y,  2020Ap&SS.365...40L}. 
Noisy components of $\delta_v$ and $\alpha$ could be considered as multiplicative noise in the system, while $N$ is additive noise. The noises $\eta_v(t)$, $\eta_\alpha(t)$ and $\eta_\mathrm{N}(t)$ are independent of each other.

In the lack of external flows and forces, Eq.~(\ref{goveEq1}) becomes a simple harmonic oscillator with linear damping, which models the rapidly decaying regime of kink oscillations. 
In the absence of noise, Eq.~(\ref{goveEq2}) reduces to the standard Rayleigh equation. For negative $\delta-\delta_v$ and positive $\alpha$, the Rayleigh equation has an asymptotically stationary (decayless) oscillatory solution with the period $2\pi/\Omega_\mathrm{k}$ and amplitude about 
\begin{equation}
\label{infamp}
\xi_\infty = \sqrt{|4 (\delta-\delta_{v0})/(3\alpha_0 \Omega_\mathrm{k}^2)|}
\end{equation} 
\citep[e.g.][]{Rayleigh}, see Fig.~\ref{fig:fig2}(a). In the phase portrait, the stationary solution is given by a limit cycle.  \citet{2016A&A...591L...5N} suggested that this solution corresponds the decayless regime of kink oscillations, which has been confirmed by full MHD numerical simulations by \citet{2020ApJ...897L..35K}.

\section{Results}
\label{SecR}

\begin{figure}
	\includegraphics[width=\columnwidth]{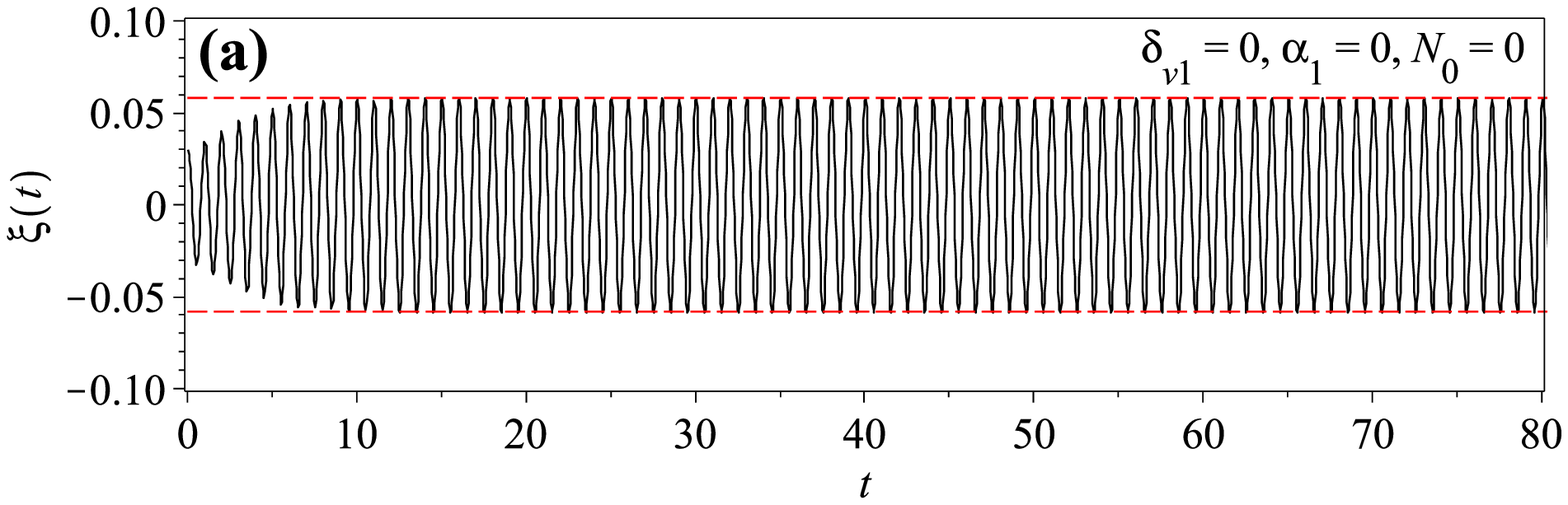}
	\includegraphics[width=\columnwidth]{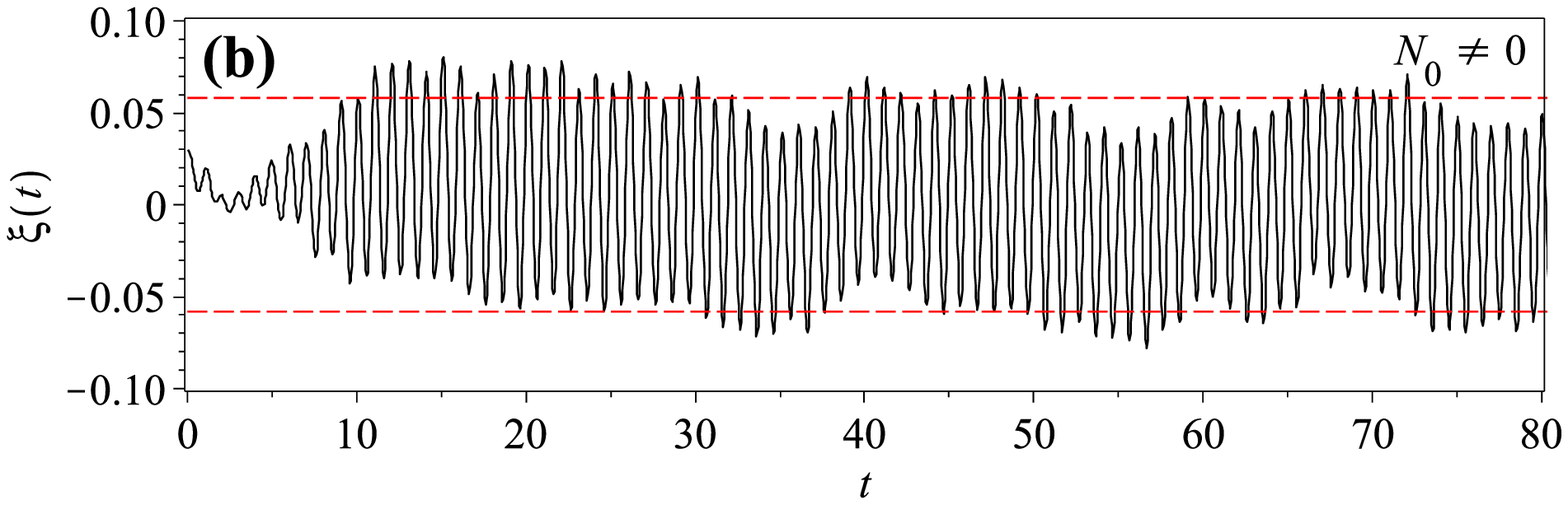}
	\includegraphics[width=\columnwidth]{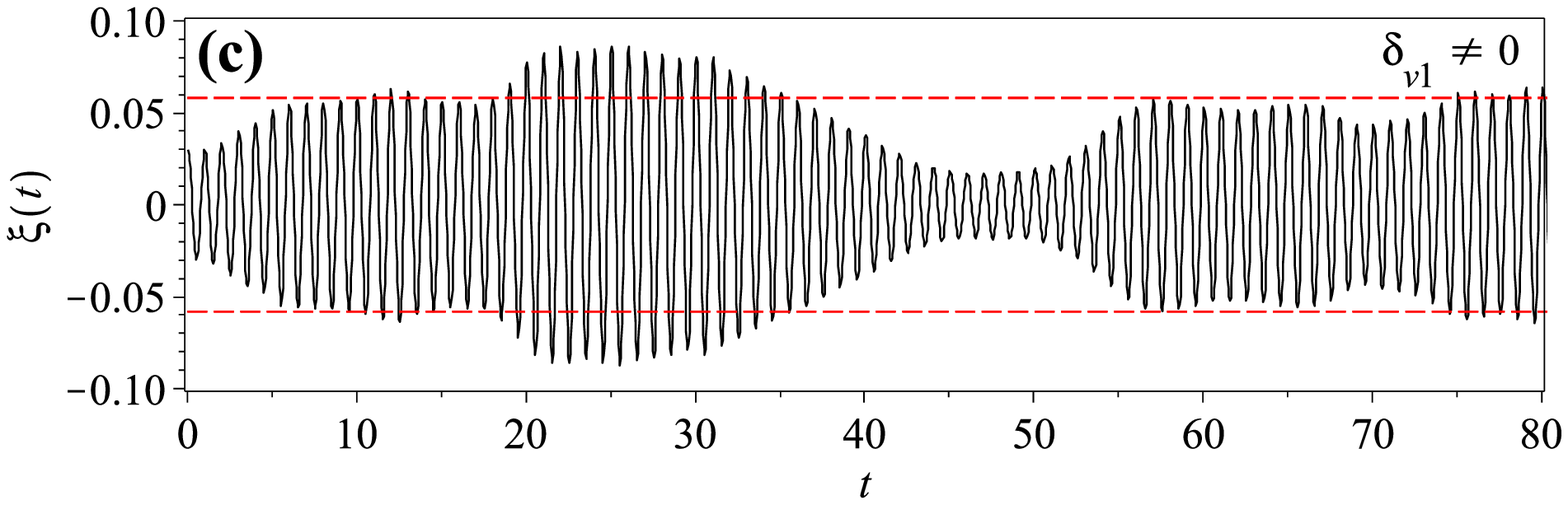}
	\includegraphics[width=\columnwidth]{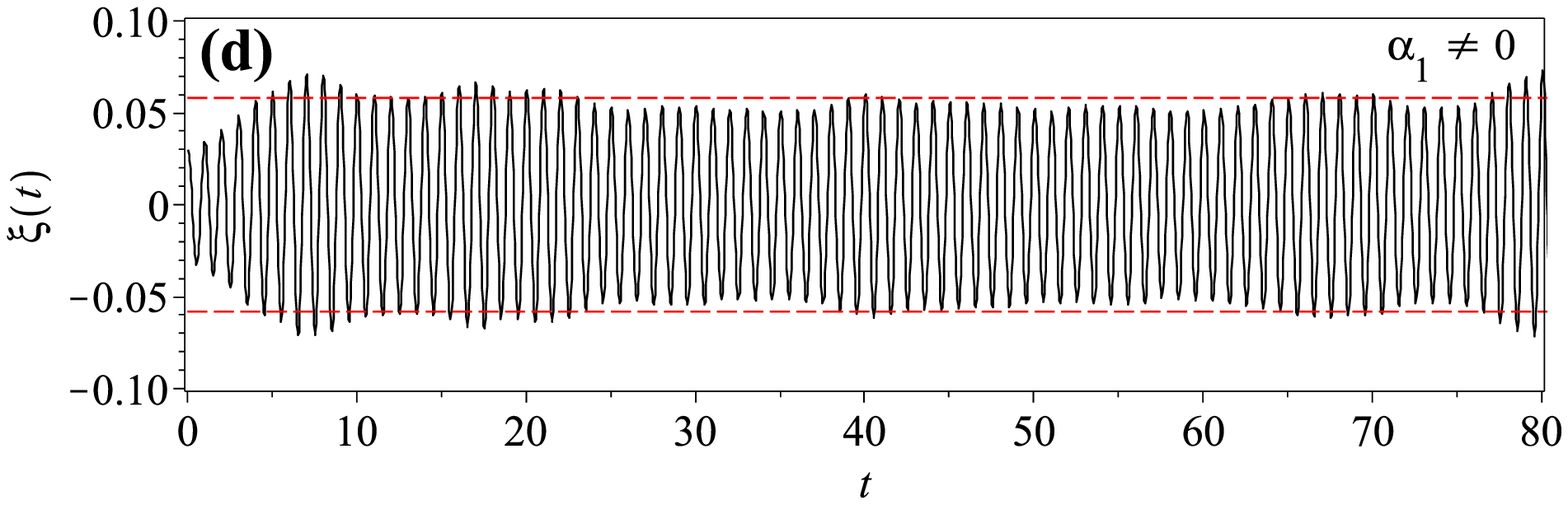}
	\includegraphics[width=\columnwidth]{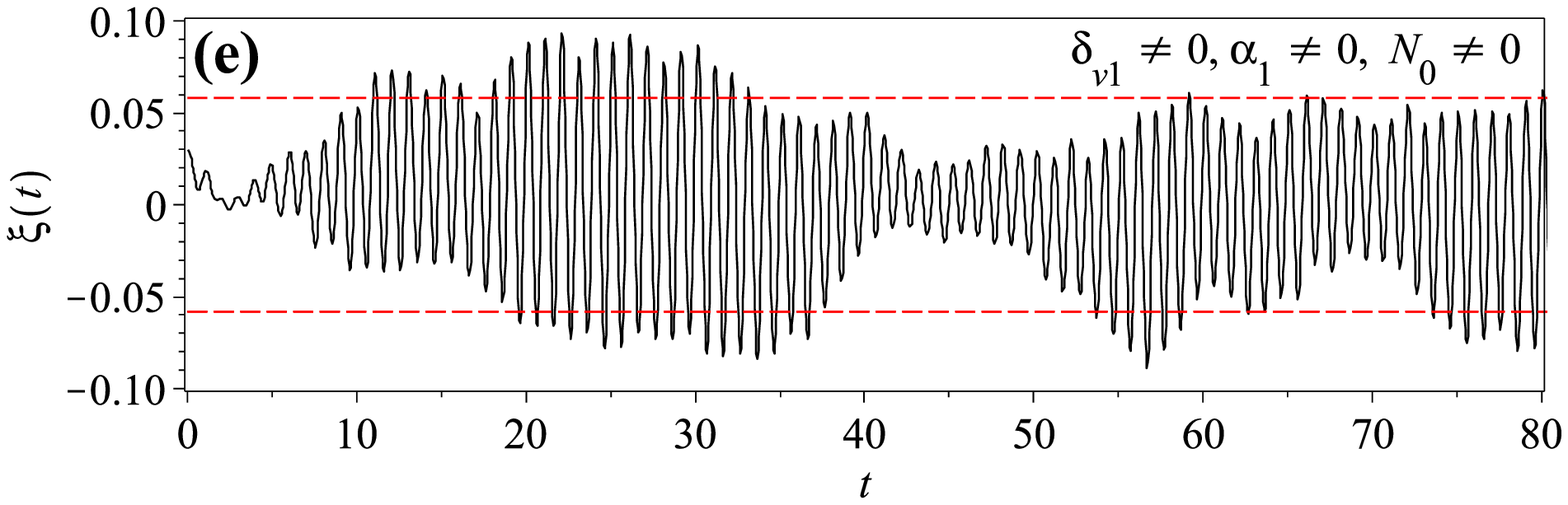}
    \caption{Typical oscillatory patterns described by a randomly-driven Rayleigh oscillator equation (\ref{goveEq2}) with random coefficients. 
    Panel (a): $\delta_{v1} = 0$, $\alpha_1=0$ and $N_0 = 0$ (no noise);
    Panel (b): $\delta_{v1} = 0$, $\alpha_1=0$ and $N_0 = 0.5$;
    Panel (c): $\delta_{v1} = 0.3$, $\alpha_1=0$ and $N_0 = 0$;
    Panel (d): $\delta_{v1} = 0$, $\alpha_1=1.2$ and $N_0 = 0$;
    Panel (e): $\delta_{v1} = 0.3$, $\alpha_1=1.2$ and $N_{0} = 0.5$. In all panels, $\delta_{v1}$ and $\alpha_1$ are measured in units of $\delta-\delta_{v0}= -0.5$ and $\alpha_0=5$, respectively. The amplitude of the additive noise $N(t)$ is measured in the units of the displacement function $\xi(t)$.
    The red dashed lines indicate the saturation amplitude $\xi_\infty$ determined by Eq.~(\ref{infamp}).
    }
    \label{fig:fig2}
\end{figure}

\begin{figure*}
    \centering
    \includegraphics[width=0.32\textwidth]{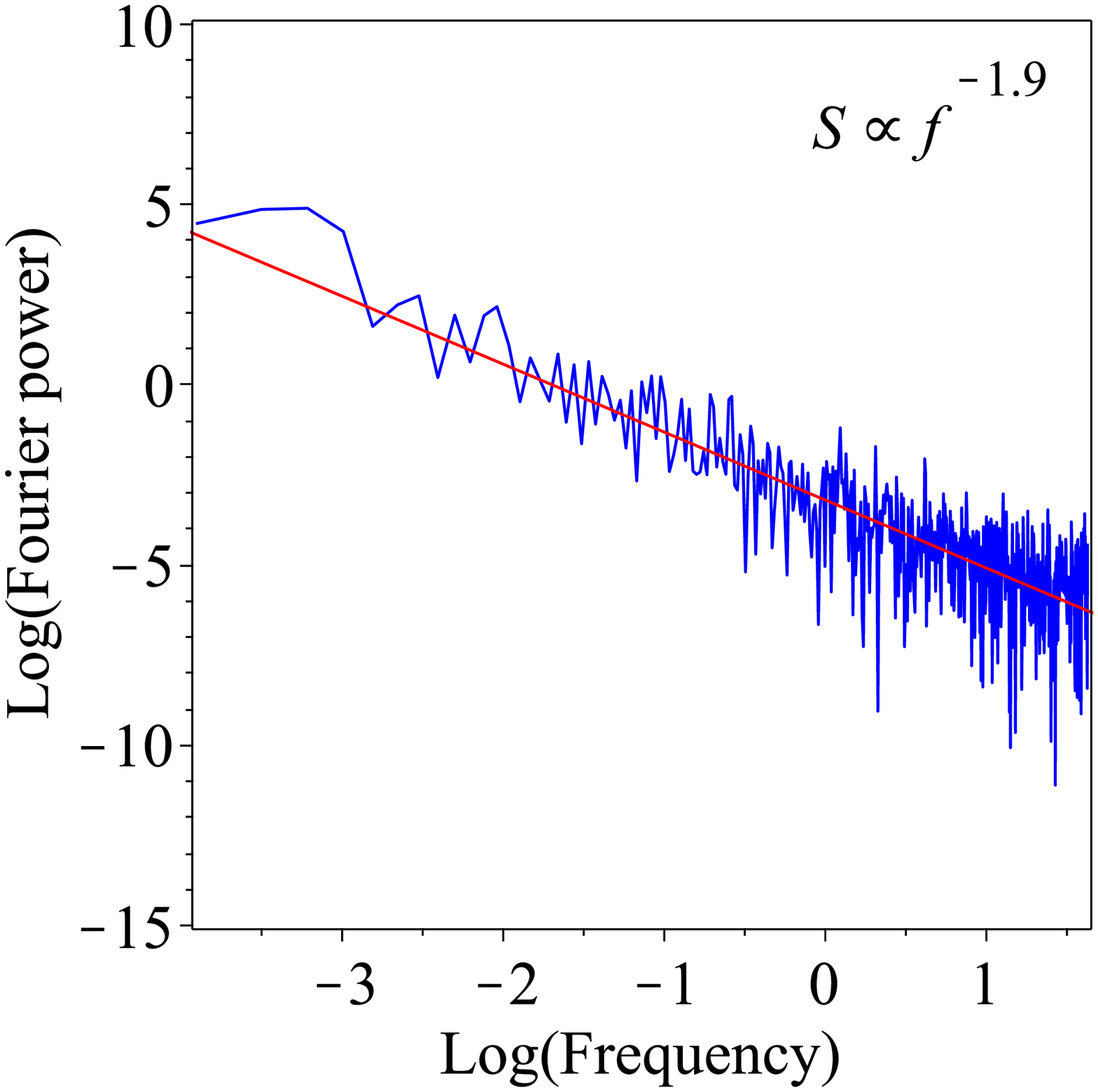}
    \includegraphics[width=0.33\textwidth]{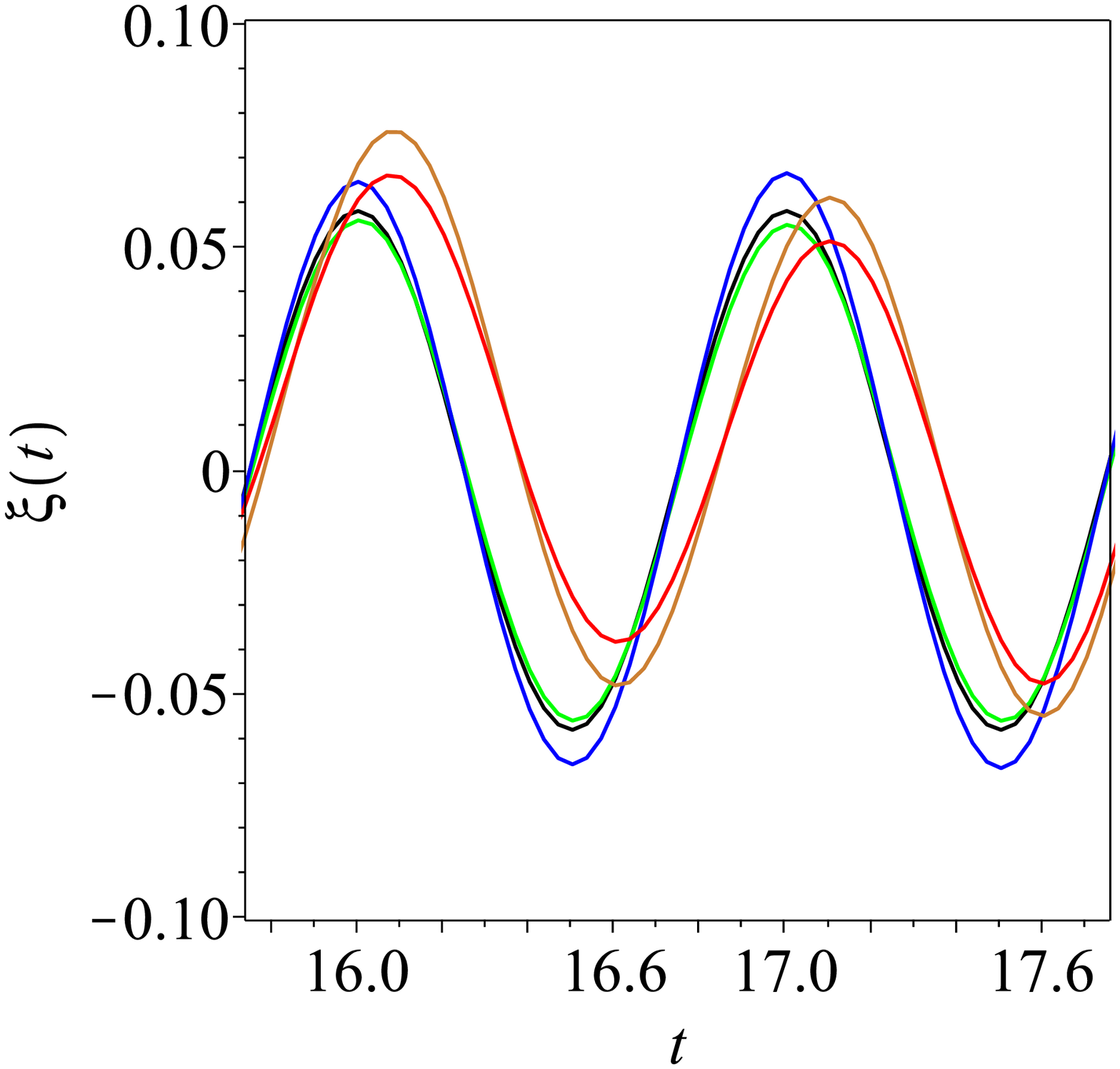}
    \includegraphics[width=0.33\textwidth]{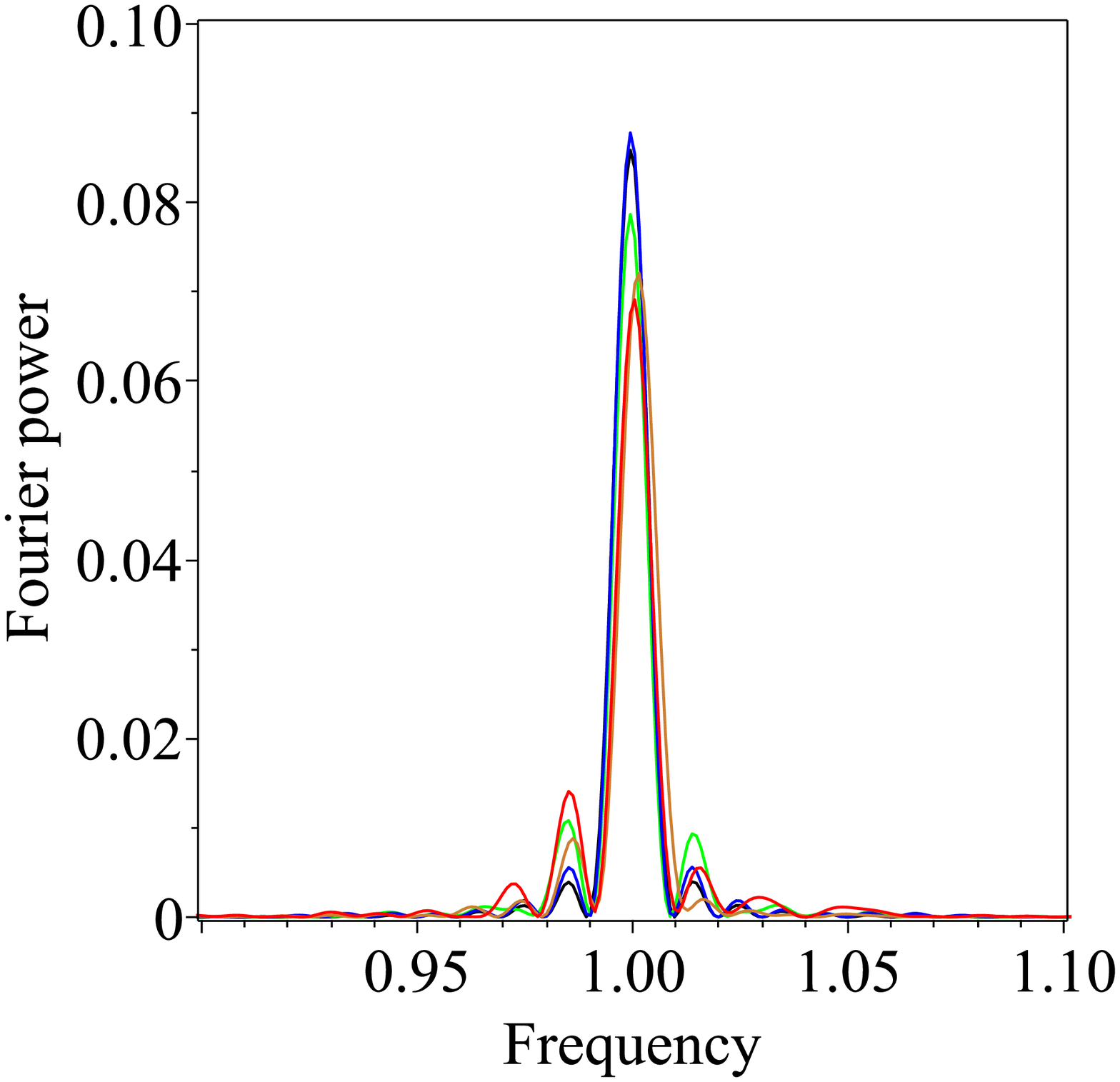}
    \caption{Left: Example of the red noise Fourier spectrum. In the inlet, the best-fitting dependence of the spectral power $S$ on the frequency $f$ is given. 
    Middle: Example of two cycles of the oscillations shown in panels (a)--(e) in Fig.~(\ref{fig:fig2}), in black, green, blue, gold, and red colour, respectively.
    Right: Fourier spectra of the oscillations shown in panels (a)--(e) in Fig.~(\ref{fig:fig2}), with the same colour scheme as in the middle panel.
    }
    \label{fig:fig3}
\end{figure*}

\begin{figure*}
	\includegraphics[width=0.33\textwidth]{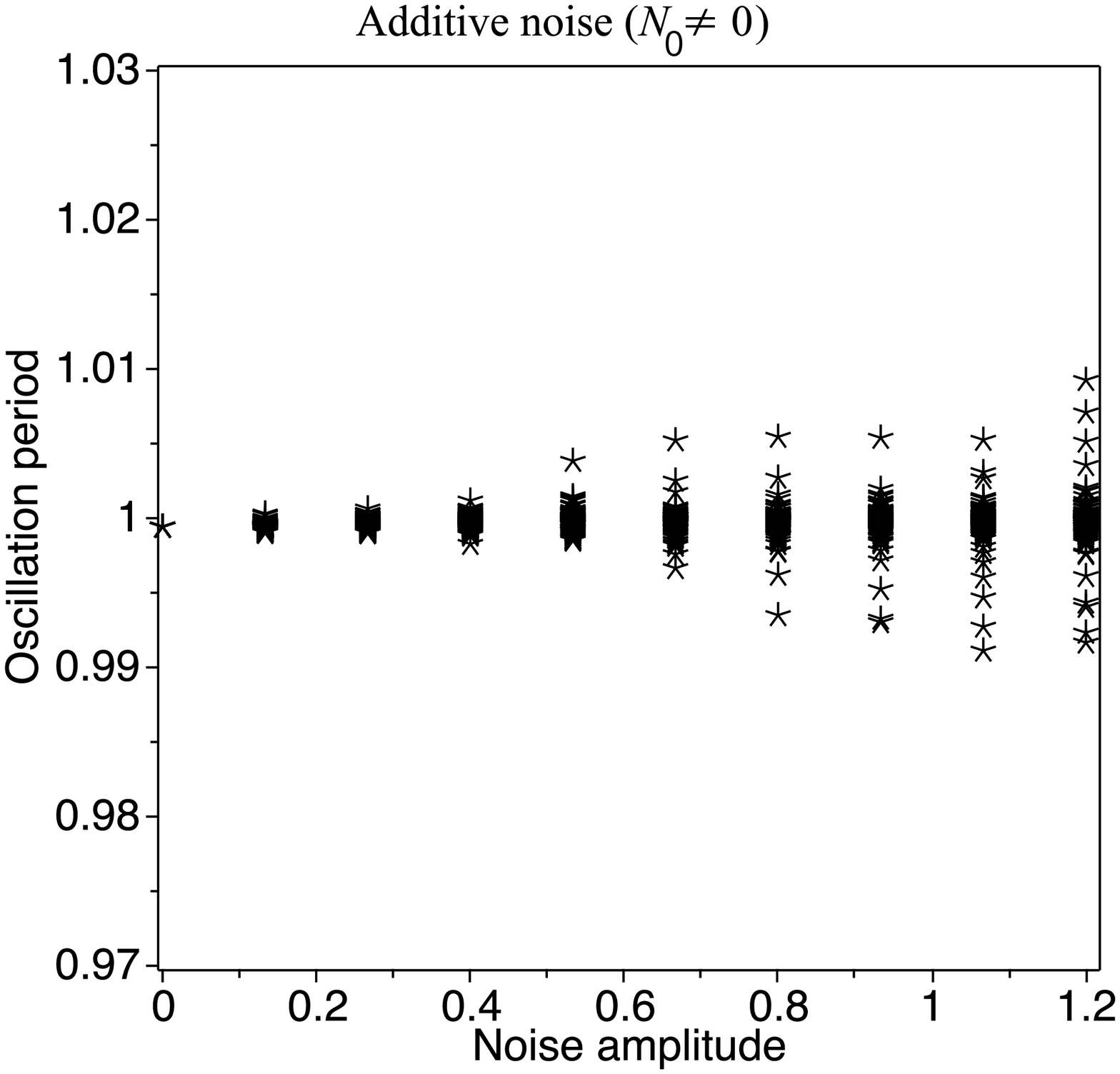}
	\includegraphics[width=0.33\textwidth]{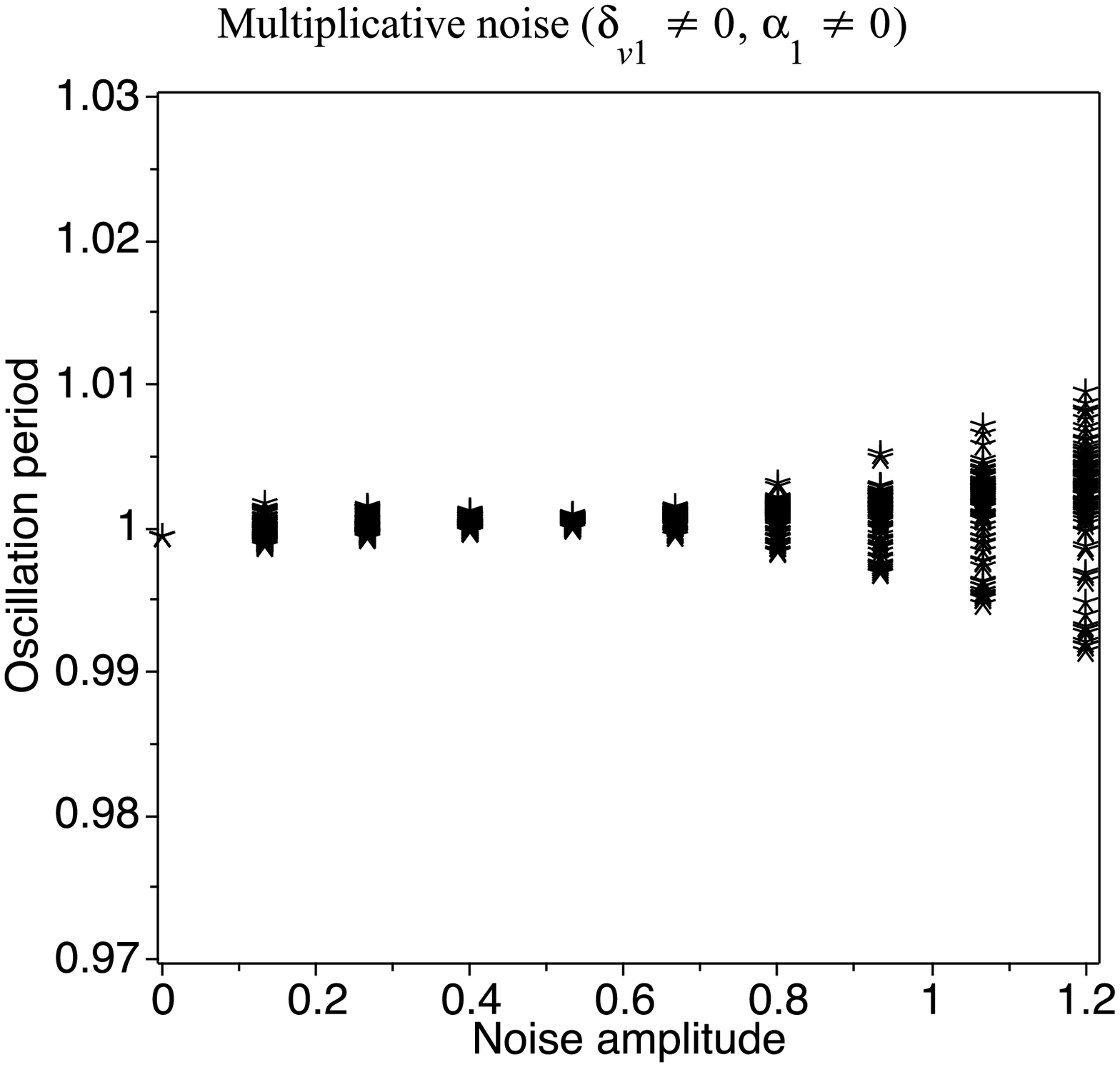}
	\includegraphics[width=0.33\textwidth]{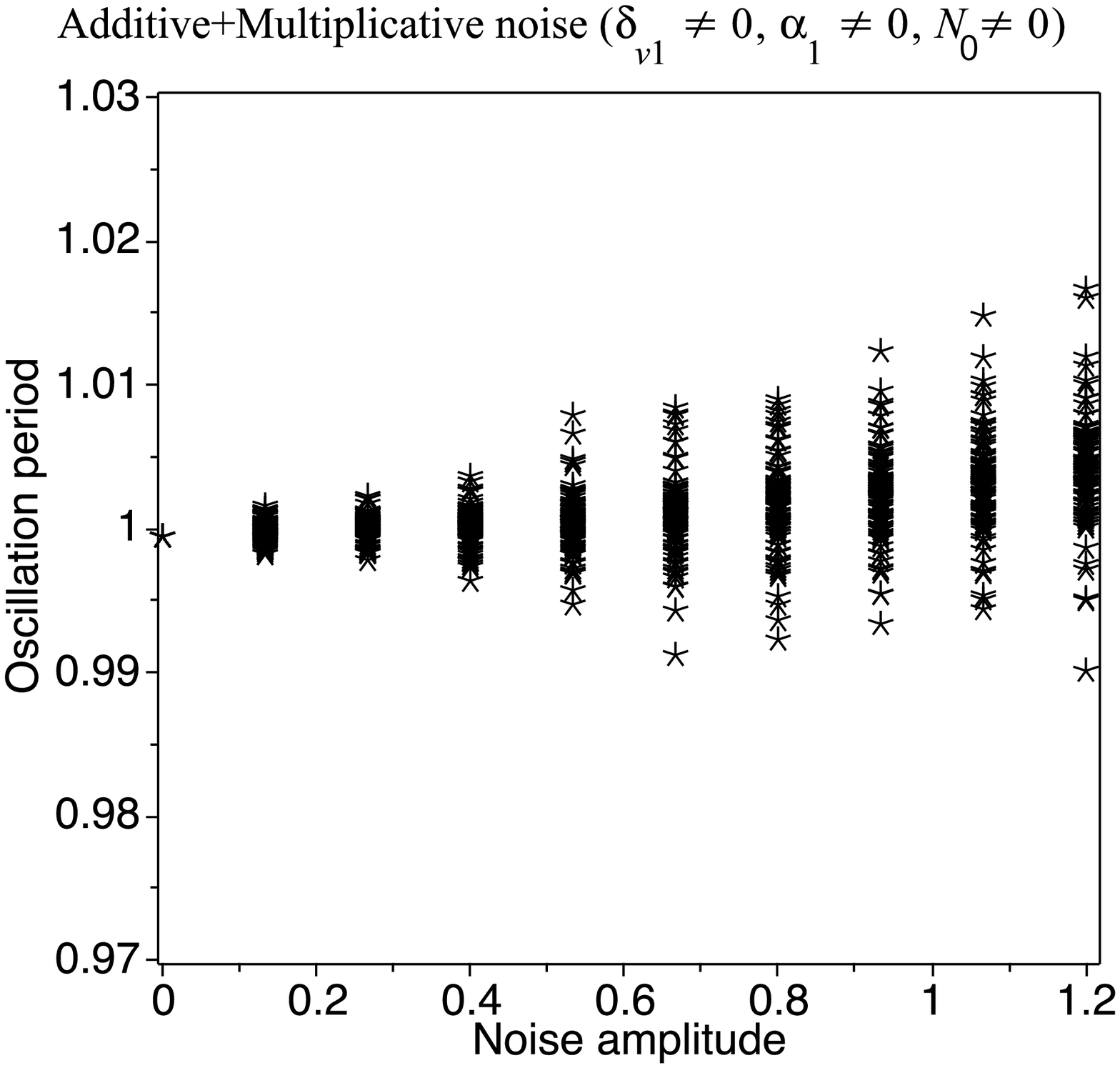}
    \caption{Dependence of the oscillation period prescribed by Eq.~(\ref{goveEq2}) on the noise amplitude for additive red noise with $N_0\neq 0$, $\delta_{v1}=0$, $\alpha_1=0$ (left), for multiplicative red noise with $N_0= 0$, $\delta_{v1} \neq 0$, $\alpha_1 \neq 0$ (middle), and for both additive and multiplicative noises with $N_0 \neq 0$, $\delta_{v1} \neq 0$, $\alpha_1 \neq 0$ (right).
    In all panels, the amplitudes of noise functions $\delta_{v1}$ and $\alpha_1$ are measured in units of $\delta-\delta_{v0}= -0.5$ and $\alpha_0=5$, respectively. The amplitude of the additive noise $N_0$ is measured in the units of the displacement function $\xi(t)$.
    }
    \label{fig:fig4}
\end{figure*}

An initial value problem constituted by Eq.~(\ref{goveEq2}) supplemented by certain initial conditions was numerically solved with the use of the \textit{dsolve} function of the \textit{Maple 2020.2} environment.
The red noise realisations were generated with the function \textit{BrownianMotion} of the \textit{Finance} package in \textit{Maple}.
Figure~\ref{fig:fig3}~left shows a typical Fourier power spectrum of the noise.

Figure~\ref{fig:fig2}(b)-(d) show typical solutions obtained when only one of the functions $\delta_{v}$, $\alpha$ and $N$ was random, and panel (e) shows the case when all those functions fluctuate randomly.  In all the cases the initial amplitude is lower than the stationary solution, hence we see the initial amplitude growth followed by random fluctuations around the stationary value $\xi_\infty$ (see Eq.~(\ref{infamp})). In the cases (not shown here) when the initial amplitude is higher than the stationary value $\xi_\infty$, the oscillation gradually approaches the stationary amplitude from above. In other words, similarly to the noiseless case, independently of the initial amplitude, the amplitude tends to the stationary value. However, in the presence of noise, the amplitude experiences fluctuations around that value. The fluctuations have a wave train nature. The amplitude fluctuations caused by random fluctuations of $\delta_{v1}$ and $\alpha_1$ on one hand, and $N$ on the other are different. In the former case, the oscillation pattern obtains a symmetric modulation, while in the latter one the modulation is rather antisymmetric. In the case when all three kinds of the noises are present, the modulation is a superposition of the symmetric and antisymmetric modulations. 

Figure~\ref{fig:fig3}~middle shows the closeup 
oscillatory patterns. 
In all cases the shape is almost harmonic. The almost monochromatic nature of the oscillations is also confirmed by the Fourier power spectra of the oscillatory signals shown in the right panel. The peak frequencies have the value about unity which corresponds to $\Omega_\mathrm{k}/2\pi$, i.e., the natural oscillation frequency. 
Figure~\ref{fig:fig4} shows the dependence of the oscillation frequency on the noise amplitudes $\delta_{v1}$, $\alpha_1$ and $N_0$, based on the use of 100 noise realisations for each value of the amplitude. The modification of the oscillation period is less than two percents in all considered cases. Thus, practically, the noise does not affect the oscillation period at all. 

\section{Discussion and Conclusions}
\label{SecD}

We studied the effect of multiplicative and additive noises on the oscillation period of decayless kink oscillations modelled by a driven Rayleigh oscillator equation. This model combines the interpretations of kink oscillations as self-oscillations generated by the interaction of a coronal loop with an external quasi-steady flow, proposed by \citet{2016A&A...591L...5N} and as oscillations driven by random motions of the loop footpoints \citep{2020A&A...633L...8A}. In the absence of footpoint movements and external coronal flows, the period of kink oscillations is prescribed by the ratio of the loop length and the kink speed, i.e., the loop acts as a kink wave resonator. The coefficients describing self-oscillations are considered to have random fluctuations. Thus, footpoint movements appear as additive noise in the governing equation, while the fluctuations of the external coronal flows constitute multiplicative noise. All fluctuations are taken to be red noises. The low-dimensional modelling is clearly a heuristic one, but it allows us to get insight into basic properties of the process in question. In particular, this approach allows us to perform a comprehensive study of the phenomenon by making a large number of computational runs with different independent noise realisations and with a weak effect of computational artefacts.

The key result is the practical independence of the kink oscillation period of noise. For both multiplicative and additive noises of all considered amplitudes the oscillation period did not differ from the natural mode period, i.e. from the value $2\pi/\Omega_\mathrm{k}$ for more than two percents. This finding justifies the seismological estimations of the kink and Alfv\'en speeds and the magnetic field in an oscillating loop by kink oscillations, based on the observed oscillation period \citep[e.g.][]{2001A&A...372L..53N, 2019ApJ...884L..40A,2020ApJ...894L..23M}. Thus, the value of the oscillation period used in the seismological formula should simply be the average of the instantaneous values. The random variations of the instantaneous oscillation period demonstrated by  \citet{2022MNRAS.513.1834Z} should be attributed to an observational effect caused by the low, sub-resolution amplitude. Obviously, this conclusion is valid only if the instantaneous periods do not show a systematic trend which could be produced by gradual evolution of the oscillating loop \citep[see, e.g.][]{2011SoPh..271...41R, 2019ApJ...875...33H}. 

The shape of the oscillatory patterns remains almost harmonic for all considered noises. In observations, it is difficult to assess the shape of the decayless kink oscillations, because the oscillation amplitude is smaller than a pixel. But, well-resolved decaying kink oscillations are seen to have almost harmonic amplitude in the low-amplitude regime. 
In the considered model, the noise causes amplitude modulation of the oscillation in a form of oscillation trains. 
This is consistent with the typical behaviour of decayless kink oscillation, in particular, the time intervals of the amplitude growth, pointed out by \citet{2012ApJ...751L..27W}, see also Fig.~\ref{fig:fig1}.
Multiplicative noise produces symmetric amplitude modulation, while the additive noise modulates the amplitude antisymmetrically, i.e., producing wiggles of the oscillation pattern. The antisymmetric wiggling trend resembles the long-period oscillatory component detected by  \citet{2016A&A...589A.136P}. 

Our model considers a single mode of the oscillations only, which corresponds, for example, to one of the kink harmonics. On the other hand, \citet{2020A&A...633L...8A} demonstrated that random motions of footpoints could excite several different parallel harmonics. Moreover, the simultaneous presence of the fundamental and second kink harmonics has been observed by \citet{2018ApJ...854L...5D}. As low amplitudes of higher harmonics make their non-linear coupling with other harmonics not likely, the developed model could be readily generalised on the case of multiple non-interacting harmonics. In addition, in our modelling we neglected effects of nonlinear damping \citep[e.g.][]{2021ApJ...910...58V}, and also the possibility of exponential and Gaussian decaying regimes \citep[e.g.][]{2013A&A...551A..40P}, which could be accounted for in a follow up study. 

\section*{Acknowledgements}
V.M.N. and D.Y.K. acknowledge support from the STFC Consolidated Grant ST/T000252/1.
S.Z. is supported by the China Scholarship Council--University of Warwick joint scholarship. 
\section*{Data Availability} 
The data underlying this article are available in the article and in the references therein.

\bibliographystyle{mnras}
\bibliography{decaylesstheory} \bsp	
\label{lastpage}
\end{document}